\documentstyle[aps,twocolumn]{revtex} 
\begin{document}
\newcommand{\bdm}{\begin{displaymath}}
\newcommand{\edm}{\end{displaymath}}
\newcommand{\be}{\begin{equation}}
\newcommand{\ee}{\end{equation}}
\newcommand{\bea}{\begin{eqnarray}}
\newcommand{\eea}{\end{eqnarray}}

\wideabs{
\title{Exact Solution of Return Hysteresis Loops in One Dimensional Random
Field Ising Model at Zero Temperature} \author{Prabodh Shukla \\ Physics
Department, North Eastern Hill University \\ Shillong-793 022, India}
\maketitle
\begin{abstract}

Minor hysteresis loops within the main loop are obtained analytically and
exactly in the one-dimensional ferromagnetic random field Ising-model at
zero temperature.  Numerical simulations of the model show excellent
agreement with the analytical results.


\end{abstract}

}

\section{Introduction} 

Hysteresis is observed in any material which is driven by a cyclic force
faster than it can equilibrate. It has practical importance, and old
scientific interest ~\cite{old} renewed by present focus of statistical
mechanics on nonequilibrium phenomena. There have been many theoretical
studies of hysteresis recently, and also simulations and experiments
~\cite{dhar,sethna,stanley}. So far only exact calculations of hysteresis
are limited to the random field Ising model (RFIM), in one dimension and
on a Bethe lattice, at zero temperature, when the driving field changes
infinitely slowly, and the system evolves from a saturated state. These
limitations are forced by our analytical abilities, but make reasonable
simplifications of physical systems in a regime where temperature effects
on hysteresis are small. The ferromagnetic RFIM model with single spin
flip dynamics at zero temperature has been proposed as a model of the
Barkhausen noise by Sethna et al ~\cite{sethna}. It covers other phenomena
as well ~\cite{other} including athermal martensitic transformations,
fluid flow in porous media, and pinning of flux lines in superconductors. 
The difficulty (in one dimension!) of an exact solution of this model lies
in the analytical treatment of quenched disorder. Even at a mean field
level, the analysis of quenched disorder can involve a formalism (e.g. 
replica method) which belies the transparency of numerical simulations. We
have used probabilistic methods to solve the antiferromagnetic RFIM in one
dimension ~\cite{psaf}, and the ferromagnetic RFIM on a Bethe lattice as
well ~\cite{shukla} . As indicated above, these solutions were restricted
to the case where the system evolves from an initial state with all spins
parallel to each other. In the present paper we are able to lift this
restriction for the ferromagnetic RFIM in one dimension. We present exact
solutions of return hysteresis loops starting anywhere on the parent loop. 
This brings the probabilistic method of solution to a level of maturity
where its application to other problems appears plausible. 

\section{Starting with a saturated state}

The one dimensional random field Ising model is characterized by the
Hamiltonian,

\be H=-J \sum_{i} s_{i} s_{i+1}- \sum_{i} h_{i} s_{i} -h \sum_{i} s_{i}
\ee

Here $s_{i}=\pm 1$ are the Ising spins, $h_{i}$ is the quenched random
field drawn from a continuous probability distribution $p(h_{i})$, and h
is the external field. The zero temperature dynamics amounts to flipping a
spin only if it lowers the energy of the system. It normally causes an
avalanche, i.e. a large number of neighboring spins have to be flipped
before the system comes to a stable state. We keep the applied field fixed
during an avalanche, and raise it afterwards until the next avalanche
occurs.

The ferromagnetic RFIM ($J \ge 0$) has two important properties. It is
abelian, i.e. the stable state after an avalanche does not depend upon the
order in which the spins flip during an avalanche. And it has return point
memory, i.e. the stable state in a slowly changing field $h$ depends only
on the state where this field was last reversed. In the special case when
we start at $h=-\infty$, and raise the field monotonically, the state at
$h$ does not depend on the rate of increase in $h$. Large rates of
increase result in fewer but larger avalanches, and small rates in more
numerous but smaller avalanches. The final state remains the same. We
exploit this property in determining the stable state at $h$ through a
single large avalanche from the initial state at $h=-\infty$. The abelian
property tells us that during this avalanche, whether a spin at site $i$
flips or not depends on the quenched field $h_{i}$ on the site and the
number of nearest neighbors n ($n=0,1,2$) which have flipped up
{\em{before}} it, but not on the order in which the neighbors flipped.
This probability is given by,

\bdm p_{n}(h)= \mbox{ prob}[h_{i}+2(n-1)J+h] \ge 0
=\int_{2(1-n)J}^{\infty} p(h_{i})dh_{i} \edm

We now need to calculate the probability that a nearest neighbor of a site
i flips up before site i.  Let us denote the conditional probability that
site i+1 (or site i-1) flips up before site i by $P^{*}(h)$. There are
many ways in which the site i+1 could be up, and we must sum over all the
possibilities to calculate $P^{*}(h)$. If site i is down, and site i+1 is
up, a spin at site i+m ($ m \ge 1$) must have flipped up before any of its
neighbors were up, and then the spins from i+m to i+1 must have flipped
up. Summing over these cases, we get

\bdm P^{*}(h) = \frac{p_{0}(h)}{1 -[p_{1}(h)-p_{0}(h)]} \edm

The probability than an arbitrary site is up at field h is given by,

\bea p(h)=[P^{*}(h)]^{2} p_{2}(h)  
+ 2 P^{*}(h)[1 - P^{*}(h)] p_{1}(h) & \nonumber \\ 
+ [1 - P^{*}(h)]^{2}p_{0}(h) & \eea

The magnetization per spin m(h) is related to p(h) by the simple equation
$m(h)=2 p(h) -1$. The lower half of the large hysteresis loop in Figure 1
shows $m(h)$ for a Gaussian distribution of the quenched field, and in
Figure 2 for a rectangular distribution. The upper half of the main loop
in each case has been obtained by symmetry $m_{u}(h)=-m(-h)$. 

\section{Reversing the applied field} Reversing the applied field from
$h=+\infty$ does not constitute a new problem because the upper half of
the large hysteresis loop shown in Figure 1 can be obtained from the lower
half by symmetry. However, reversing the applied field from any other
point constitutes a new and somewhat more difficult problem. The reason is
that in a starting state at a finite field $h$, whether the spin at a site
is flipped or not depends in a nontrivial way on the random field at that
site as well as on neighboring sites. The state is thus "strongly
correlated", and it is difficult to do perturbation theory about this
state.

For an arbitrary starting state on the lower hysteresis loop, the spins
which can initiate a downward avalanche have to be separated into at least
nine different categories; three categories depending on the number of
nearest neighbors which are up in the starting state (n=0,1,2; these are
the number of up neighbors of an up spin {\em{after}} the upward avalanche
has settled at the point of return), and three categories depending on the
number of up neighbors just {\em{before}} it flips up during an avalanche.
The number of up neighbors {\em{before}} turning up in an avalanche
remains important even after the avalanche because it determines the
{\it{a posteriori}} distribution of the quenched field on the up spins in
the stable state at the point of return. We also use three more categories
characterized by the number of up neighbors just before a spin turns down
in a downward avalanche (this number is different from the one at the
starting point of the reverse trajectory).

We start backtracking from an arbitrary applied field $h$ on the lower
loop, and come down to $h^{\prime}$ ( $h^{\prime} \le h$). We want the
magnetization at $h^{\prime}$.  Obviously, spins can only flip down on the
reverse trajectory, and therefore we focus on spins which are up at $h$
but turn down at $h^{\prime}$. We divide the up spins at $h$ into three
basic categories characterizing the range of their random field, and how
they turned up on the lower hysteresis loop. Spins in category-0 have
$h_{i} \ge 2J-h$. These spins could turn up at $h$ even if none of their
neighbors were up to help them. Spins in category-1 have $-h \le h_{i} \le
2J-h$, and spins in category-2 have $-2J-h \le h_{i} \le -h$. No spin
could be up at $h$ if it has $h_{i} \le -2J-h$. How the spins turn down at
$h^{\prime}$ on the reverse trajectory is determined by the random field
on a spin and the number of up neighbors it has just before it turns down.
The three basic categories listed above were determined by the number of
up neighbors just before a spin turned up during an upward avalanche at
$h$. After that avalanche has settled, the number of up neighbors may
increase. Thus each of the basic categories can be further divided into
three categories characterized by the number of up neighbors after the
avalanche. Some of these sub-categories may be empty. For example, a spin
of category-2 which is up at $h$ necessarily has both neighbors up. Spins
of category-1 could have one or both neighbors up. Spins of category-0
could have zero, one, or two neighbors up at $h$.  When the applied field
is reversed, spins of category-2 with both neighbors up are as susceptible
to turn down as spins of category-1 with one neighbor up because the net
field in both cases lies in the same range.

In the first instance, we consider a restricted range of the reversed
field: $h-2J \le h^{\prime} \le h$. In this range, the only spins which
could turn down are spins of category-2 with two neighbors up, spins of
category-1 with one neighbor up, and spins of category-0 with zero
neighbors up. We add the contributions from these three categories, and
subtract the sum from the number of up spins at $h$. This gives us the
magnetization at $h^{\prime}$. Consider the spins of category-2 first;
their fraction at $h$ is equal to $[P^{*}(h)]^{2} [p_{2}(h) - p_{1}(h)]$.
The factor $[P^{*}(h)]$ gives the probability that a nearest neighbor of a
spin is up on the lower hysteresis loop before that spin is relaxed. Thus
$[P^{*}(h)]^{2}$ is the probability that both neighbors of the spin are up
before it is relaxed. The factor $[p_{2}(h) -p_{1}(h)]$ gives the
probability that the spin turns up if two neighbors are up but not if only
one neighbor is up. Thus, the fraction of category-2 spins which turn down
at $h^{\prime}$ on the return loop is given by,

\bdm q_{r}^{2}(h,h^{\prime}) =[P^{*}(h)]^{2}[p_{2}(h) -
p_{2}(h^{\prime})]. \edm

Now we take up the spins of category-1. In the initial state at $h$,
category-1 spins come in two sub-categories; (i) spins with one neighbor
up, and (ii) spins with two neighbors up. In the restricted range of the
reversed field ($h-2J \le h^{\prime} \le h$), spins in sub-category (ii)
can not turn down spontaneously. However they can turn down in an
avalanche, if the avalanche puts one of their neighbors in category (ii)
and it turns down. An avalanche can start with a category-1 spin which has
one neighbor down in the starting state at $h$. This occurs with the
probability $f(h)$ given by,

\bdm f(h) = \{1-p_{2}(h)\} [P^{*}(h)] + \{1-p_{1}(h)\} [1-P^{*}(h)] \edm

The above equation can be understood as follows. Suppose, the spin at site
i is up, and $f(h)$ denotes the probability that the spin at site i+1 is
down. Before the spin at site i+1 is relaxed, the spin at site i+2 is up
with the probability $[P^{*}(h)]$, and down with the probability $[1 -
P^{*}(h)]$. The probability that the spin stays down in the two cases even
after it is relaxed is given by $\{1-p_{2}(h)\}$ and $\{1-p_{1}(h)\}$
respectively. The probability for the spin at i+1 to flip down at
$h^{\prime}$ is equal to $[p_{1}(h) - p_{1}(h^{\prime})]$. After it flips
down, the spin at i-1 can also flip down with the same probability if it
belongs to category-1 and the spin at i-2 is up. Thus an avalanche can
start. The avalanche will go on till it meets a category-1 spin which does
not flip down at $h^{\prime}$ or it meets a category-0 spin which has an
up neighbor on the other side. The probability that a nearest neighbor of
an up spin is down at $h^{\prime}$ is given by,

\bdm q_{a}(h,h^{\prime}) = \frac{f(h)}{1 - [p_{1}(h) - p_{1}(h^{\prime})]}
\edm

Here, $f(h)$ is the probability that the neighbor was already down in the
initial state. The other factor is the sum of an infinite series which
accounts for avalanches of various sizes which may bring the neighbor
down.

An avalanche can also be started by a spin of category-2 flipping
down. This gives another term,

\bdm q_{b}(h,h^{\prime}) = \frac{[p_{2}(h) - p_{2}(h^{\prime})]
[P^{*}(h)]} {1 - [p_{1}(h) - p_{1}(h^{\prime})]} \edm

The numerator in the above equation can be understood as follows. Suppose
the spins at sites i, i+1, and i+2 are up and site i+1 belongs to
category-2. $[P^{*}(h)]$ is the probability that site i+2 was up before
site i+1 was relaxed at $h$. The numerator gives the probability that the
right side neighbor of the up spin at site i flips down at $h^{\prime}$.
The denominator takes care of any possible avalanches started by the
flipping down of a category-2 site. The total probability that a nearest
neighbor of an up spin is down at $h^{\prime}$ is equal to $q_{a} +
q_{b}$. We also need the probability that a nearest neighbor of an up spin
is up before that spin is relaxed at $h^{\prime}$. This is equal to the
probability that the neighbor in question was up on the lower hysteresis
loop before the site was relaxed at $h$, i.e. it is equal to $P^{*}(h)$.
With this knowledge, we are now in a position to write the fraction of
category-1 spins which turn down on the return loop at $h^{\prime}$. We
get,

\bdm q_{r}^{1}(h,h^{\prime}) =2 [P^{*}(h)] [q^{a}(h,h^{\prime}) +
q_{b}(h,h^{\prime})] [p_{1}(h) - p_{1}(h^{\prime})] \edm

Spins of category-1 can not have both nearest neighbors down. The reason
is that this class of spins are flipped up during an avalanche on the
lower hysteresis loop. Therefore they must be connected by up spins to a
spin of category-0 on one side at least. A spin of category-0 can not turn
down if it has at least one neighbor up. However, if both neighbors of a
spin of category-0 are down at $h^{\prime}$, it may turn down. The
fraction of such spins is given by,

\bdm q_{r}^{0}(h,h^{\prime}) =[q^{a}(h,h^{\prime}) +
q_{b}(h,h^{\prime})]^{2} [p_{0}(h) - p_{0}(h^{\prime})] \edm

We are now in a position to write the magnetization on the return loop in
range $[h-2J \le h^{\prime} \le h]$. We get, $m^{\prime}(h^{\prime}) = 2
p^{\prime}(h^{\prime}) - 1$, where

\be
p^{\prime}(h^{\prime}) = p(h)
-q_{r}^{2}(h,h^{\prime}) 
-q_{r}^{1}(h,h^{\prime}) 
-q_{r}^{0}(h,h^{\prime}) 
\ee

The key to getting the return magnetization beyond the range considered
above is to note that the state of the system on the reverse trajectory at
$h^{\prime}=h-2J$ is the same as would be obtained from the initial state
at $h^{\prime}=+\infty$. If the initial state at $h^{\prime}=\infty$ is
exposed to an applied field $h-2J$, spins with $h_{i} \le -h$ will flip
down spontaneously and start avalanches where the adjacent spins in the
range $-h \le h_{i} \le 2J-h$ will flip down. When this avalanche is
finished, the remaining up spins will belong to three categories: (i) spins
with $h_{i} \ge 2J-h$ with one neighbor up, (ii) spins with $h_{i} \ge
4J-h$ with no neighbors up, and (iii) spins with $h_{i} \ge -h$ with two
neighbors up. This is precisely the state obtained at the end of the
reverse trajectory obtained above. Therefore, the reverse trajectory in
the range $h^{\prime} \le h-2J$ merges with the upper half of the big
hysteresis loop.

\section{Reversing the field again} 
The magnetization in reversed field merges with the upper half of the big
hysteresis loop when the field falls below $h-2J$. Pulling up the field
from below $h-2J$ can be related by symmetry to the problem of the return
loop analyzed in the previous section. We need not repeat this
calculation. However, if the reversed field is re-reversed before it
reaches $h-2J$, we have a new problem on our hands which we now analyze.

We turn back the field at $h^{\prime}$. Our object is to obtain the
magnetization at an arbitrary value $h^{\prime \prime}$ ($h^{\prime} \le
h^{\prime\prime} \le h)$ on the lower half of the return loop.
Essentially, we are looking at the same strings of spins which turned down
in the previous section, but now they turn up from the other end. If a
spin is down on the lower half of the return loop, it must have been down
at end of the upper half as well. The reason is that on the lower half,
spins can only flip up, none can flip down. Thus the probability that a
nearest neighbor of a down spin is down on the lower return loop is equal
to $q_{a}(h,h^{\prime}) + q_{b}(h,h^{\prime})$. The probability that the
nearest neighbor is up increases steadily as more spins flip up on the
lower half. First, let us look at the probability of an up neighbor at the
start of the lower return loop. Consider three adjacent sites: i-1, i, and
i+1. Given that site i+1 is down, we want the probability that site i is
up. It follows from the previous section that if site i is up at
$h^{\prime}$, it must be a spin of category-0, or there must be a string
of up spins to the left of i containing a spin of category-0. Spins of
category-0 are up with probability unity if they are adjacent to an up
spin, otherwise they have to have a quenched field in excess of $4J-h$.
Thus the probability that site $i$ is up and is a spin of category-0 is
equal to $[1-(q_{a} + q_{b})] p_{0}(h)  + (q_{a} + q_{b})
p_{0}(h^{\prime\prime})$. The probability that site $i$ is up and not a
spin of category-0 is equal to $[P^{*}(h)] [p_{1}(h^{\prime}) -
p_{0}(h)]$.  Putting it together, the probability that a nearest neighbor
of a down spin is up on the lower return loop before that neighbor is
relaxed is given by:

\bdm p_{rr}(h,h^{\prime},h^{\prime\prime}) = \frac{a} {1 - [
p_{1}(h^{\prime\prime}) - p_{1}(h^{\prime}) ] } \edm

where, \bea a = [p_{1}(h^{\prime}) - p_{0}(h)] P^{*}(h)  + [1- ( q_{a} +
q_{b} ) ] p_{0}(h) & \nonumber \\ + (q_{a} + q_{b})
p_{0}(h^{\prime\prime}) &  \nonumber \eea

The magnetization on the lower return loop is given by
$m^{\prime\prime}(h^{\prime\prime}) = 2 p^{\prime\prime}(h^{\prime\prime})
- 1$, where

\bea p^{\prime\prime}(h^{\prime\prime}) = p^{\prime}(h^{\prime}) + (q_{a}
+ q_{b})^{2} [p_{0}(h^{\prime\prime}) - p_{0}(h^{\prime})] & \nonumber \\
+ 2(q_{a} + q_{b})  p_{rr}(h,h^{\prime},h^{\prime\prime})  
[p_{1}(h^{\prime\prime}) - p_{1}(h^{\prime})] & \nonumber \\ +
p_{rr}^{2}(h,h^{\prime},h^{\prime\prime}) [p_{2}(h^{\prime\prime}) -
p_{0}(h^{\prime})] & \eea

As may be expected, the analytical results agree quite well with numerical
simulations of the model. Figure 1 shows a comparison for a Gaussian
distribution of the random field, and Figure 2 for a rectangular
distribution. Analytical expressions are shown by continuous lines.
Simulations for a chain of $1000$ spins (averaged over $1000$ different
realizations of the random field distribution) are indistinguishable from
the analytical expressions, but these are shown by large symbols at sparse
intervals for visual convenience.

\section{Concluding remarks} 

The nonequilibrium response of RFIM at zero temperature is related to
experimentally measurable quantities in several diverse systems. It has
been calculated analytically in one dimension using probabilistic methods,
and checked against numerical simulations of the model. It remains for the
future to apply the present method in higher dimensions, although it
should be qualitatively similar.

I thank Deepak Dhar for critical reading of the manuscript.

Caption Figure 1: Hysteresis loop (filled squares) between two saturated
states for a Gaussian random field (mean=0, variance=1, J=1). Two
excursions from the lower half are shown; $h=1$ to $h^{\prime}=-1$ and
back (open squares), and $h=1$ to $h^{\prime}=-.6$ and back (open
circles).

Caption Figure 2: Hysteresis loop (filled squares) for a rectangular
distribution of the random field of width 6 (J=1). Return loop (open
squares) shows an excursion from the lower half ($h=1.5$ to
$h^{\prime}=-.5$ and back).

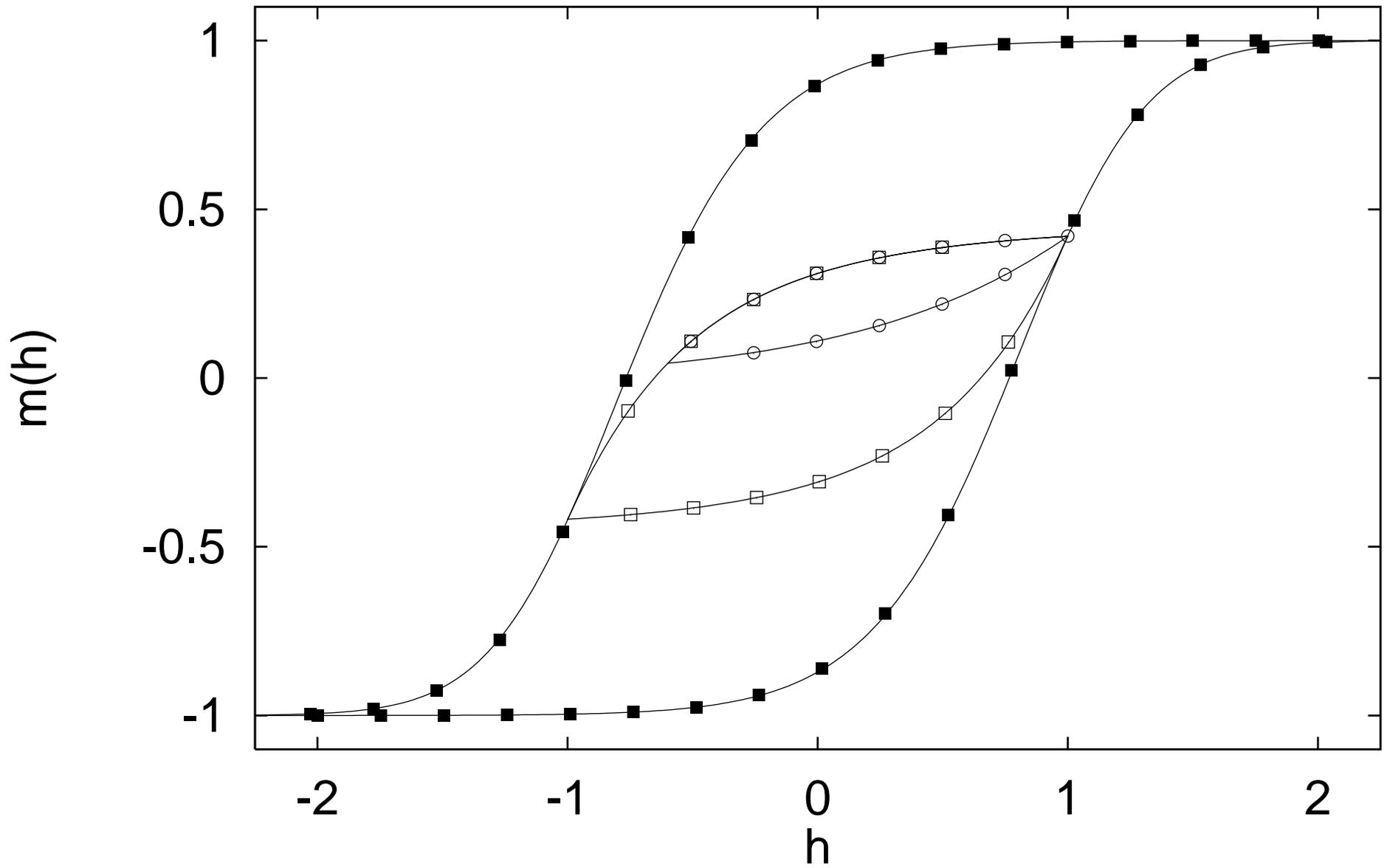
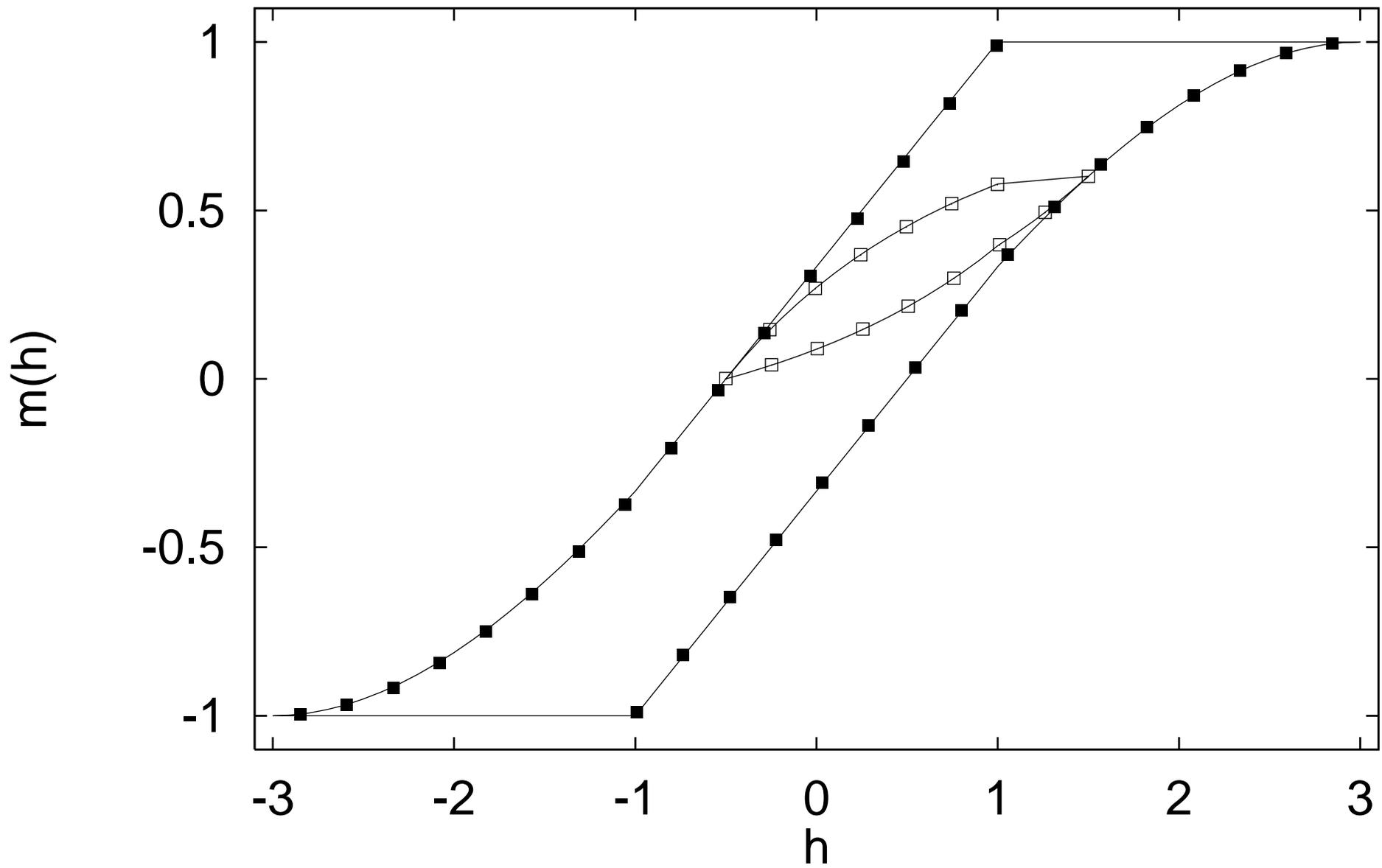
\end{document}